# Development of a 120-mm Aperture Nb$_3$Sn Dipole Coil with Stress Management

I. Novitski, A.V. Zlobin, J. Coghill, E. Barzi, *Senior Member, IEEE,* and D. Turrioni,

*Abstract*—This paper describes a 120-mm aperture 2-layer dipole coil with stress management (SM) developed at Fermilab based on cos-theta coil geometry. A model of the coil support structure made of plastic was printed using additive manufacturing technology and used for practice coil winding. The real coil support structure was printed using the 316 stainless steel. The results of the SM structure size control and the key coil fabrication steps are reported in the paper. The design of coil SM structure and the coil FEA in the dipole mirror test configurations are presented and discussed.

## I. Introduction

HIGH-field and/or large-aperture dipole and quadrupole magnets based on Nb$_3$Sn superconductor are needed for various accelerator systems of future hadron and muon colliders [1], [2]. High currents and strong magnetic fields lead to large Lorentz forces and, thus, to significant mechanical strains and stresses which can degrade or even permanently damage brittle Nb$_3$Sn coils. These forces significantly increase with the level of magnetic field and magnet aperture and at some level needs special treatment.

To provide a stable turn position and to keep the strain/stress in brittle coil within an acceptable range for the superconductor during magnet fabrication and operation, various stress management coil structures have been proposed and are being studied [3]-[5]. These structures have complex 3D geometries that are difficult to produce using traditional machining, and rather tight tolerance requirements. Recent advances with Additive Manufacturing (AM) technologies using various materials allow them to be considered for fabrication of precise and complicated metallic parts used in the magnet coils of high field accelerators. Additionally, AM is being also used for rapid prototyping of parts using inexpensive plastic materials. New methods to measure geometrical parameters of complicated 3D coil parts are also being developed.

This paper describes a 120-mm aperture 2-layer (2L) dipole coil with stress management (SM) developed at Fermilab based on cos-theta coil geometry and Nb$_3$Sn Rutherford cable. The key development and fabrication steps of the Nb$_3$Sn demo coil with SM, and the expected coil performance in a 2L and 4L dipole mirror configurations with an additional 60-mm aperture Nb$_3$Sn insert coil are reported in the paper.

This work is supported by Fermi Research Alliance, LLC, under contract No. DE-AC02-07CH11359 with the U.S. Department of Energy

Authors are with the Fermi National Accelerator Laboratory (FNAL), Batavia, IL 80510 USA (e-mail: zlobin@fnal.gov).

## II. SMCT Coil Design and Technology

### A. Coil Design with Stress Management

The cross-section of the large-aperture superconducting dipole coil developed at Fermilab using ROXIE [6] based on stress management cos-theta (SMCT) geometry is shown in Fig. 1 (center). The SMCT demo coil consists of 2-layers wound into its stainless steel structure [7]. Each coil layer is split into 5 blocks, with the number of turns in blocks approximately following the cos-theta distribution, and the equal 5-mm spacing between the blocks. The coil layers are separated by 5 mm in the radial direction from each other and from the inner coil to create space for the SMCT support structure. The support structure provides turn positioning and stress management during coil fabrication, magnet assembly and operation. Each coil block is placed in its groove and supported separately. As a result, azimuthal and radial components of the electromagnetic force applied to coil blocks are not accumulated but transmitted to the coil and magnet structures bypassing the coil blocks.

To apply the radial force to the SMCT coil structure, this coil will be tested with 60-mm aperture Nb$_3$Sn dipole insert coils developed at Fermilab for 15 T dipole demonstrator MDPCT1 [8]. Thus, the coil inner diameter is 123 mm leaving ~0.5 mm of radial space for the coil radial insulation. The coil outer diameter is 206 mm.

The design of lead and non-lead ends of the SMCT demo coil was optimized using the BEND program [9] to minimize bending stress of Rutherford cable during winding and limit the cable motion under Lorentz forces in coil ends to avoid coil quenches. The 3D views of the lead and non-lead ends of the SMCT coil outer layer are shown in Fig. 1 (left and right).

The SMCT coil uses the 15.1 mm wide and 1.319 mm thick 40-strand Rutherford cable with a keystoned cross-section used in MDPCT1 outer coils [8]. The cable is based on the Nb$_3$Sn composite wire with a Cu/nonCu ratio of 1.13 and $J_c$ at 15 T, 4.2 K of 1500 A/mm$^2$.

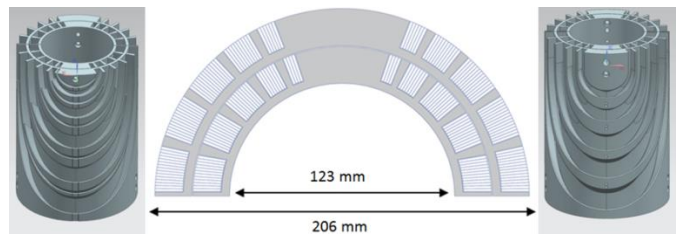

Fig. 1. Cross-section of the large-aperture 2-layer SMCT coil (center), and 3D view of the lead (left) and non-lead (right) ends of the SMCT coil structure.

## B. SMCT Practice Coil

Due to high complexity of the end geometry and large variations of the cable mechanical rigidity, the end parts of the SMCT demo coil were prototyped and tested in practice winding. The plastic parts for the SMCT practice coil are shown in Fig. 2 (left). These parts were printed at Fermilab using Stratasys F170 FDM printer and corresponding CAD models.

The coil practice winding with plastic parts allows optimizing the coil part design before procurement, checking important coil fabrication steps (except for the coil reaction), and testing tooling. In the case of SMCT coils, practice winding permitted also learning new procedure of cable placement in slots, checking the effect of cable rigidity during winding, and confirming parts sizes, cable positioning, room for cable expansion during reaction (included in the slot design based on the known cable expansion parameters), etc. Practice coil on the winding table and a 3D view of practice coil return end with axial and transverse cuts are shown in Fig. 3. The analysis of these cuts confirmed the optimal turn position and minimal gaps between the turns and structure inside coil winding.

## C. SMCT Coil Part Fabrication and Measurements

Parts for the SMCT demo coil were printed at GE Additive using L-PBF technology. SMCT coil parts made of 316 stainless steel are shown in Fig. 2 (right). Due to special requirements to the tolerances, all the coil parts were measured before coil winding. The complex 3D geometry of SMCT coil end parts with narrow slots and inclined surfaces made it difficult for traditional coordinate measuring machines (CMM), used to control dimensions of the coil straight section parts. Therefore, the end parts were measured by Laser Scanning system in free condition using ROMER Absolute Arm external scanner. Examples of surface deviations from CAD in mm obtained using CMM measurements for straight section part and Laser Scanning measurements for return end block of SMCT coil are shown in Fig. 4. Best fit of measured data gives deviations of structure inner and outer diameters within ±0.25 mm and 3D surfaces within 0.5 mm. The range of measured surfaces roughness using a stylus probe $R_a$ is within 5.5–10.5 μm.

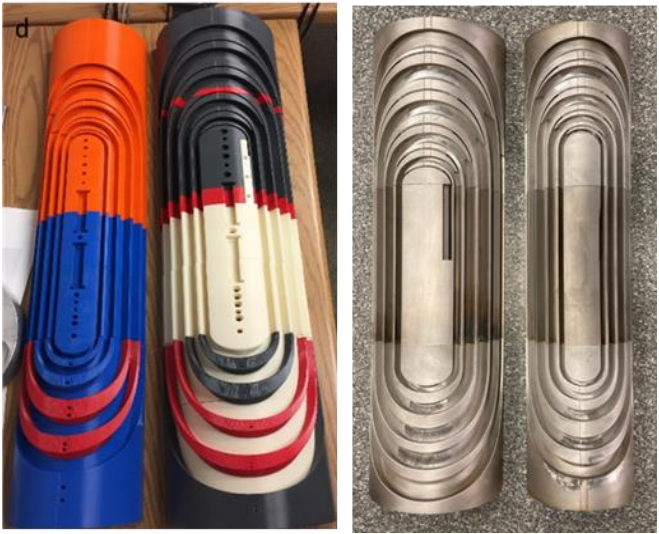

Fig. 2. Inner-layer and outer-layer plastic (left) and 316 stainless steel (right) parts for SMCT coils.

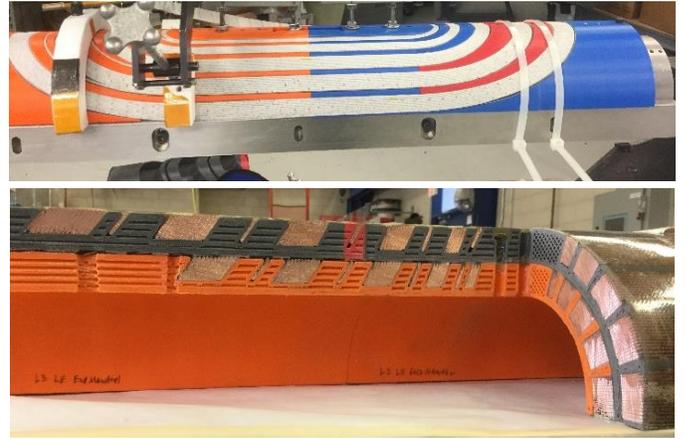

Fig. 3. Two-layer SMCT practice coil winding (top), and axial and transverse cuts of the coil return end (bottom).

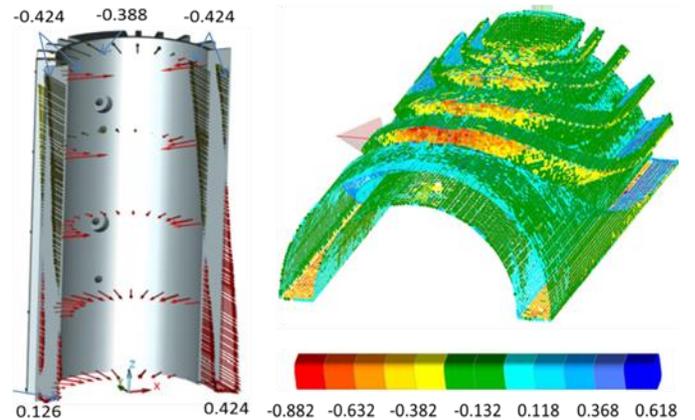

Fig. 4. Surface deviations from CAD for a straight section part of the SMCT coil measured with CMM (left) and for return end block of the SMCT coil measured with Laser Scanning (right). The numbers in the pictures are in mm.

The measurement results show that the accuracy of SMCT coil parts produced with the AM technology is still lower than the accuracy of traditional end spacers made by mechanical or electrical discharge machining (EDM). However, the accuracy of the AM parts can be considered as acceptable for the first-time 3D printing of metal objects with complicated 3D geometries and relatively high tolerance requirements. Various possibilities of improving the 3D part printing are being studied.

## III. SMCT COIL TEST IN MIRROR CONFIGURATION

### A. Test Configurations and Conductor Limit

The first large-aperture SMCT demo coil will be tested in a dipole mirror configuration using MDPCT1 mechanical structure [8]. To accommodate the SMCT coil with larger outer diameter the inner diameter of iron laminations was increased from 196.1 mm to 209.5 mm. The cross-section of the 4L dipole mirror with outer SMCT coil and the inner 60-mm aperture coil developed for the 15 T dipole MDPCT1 inside the modified MDPCT1 support structure is shown in Fig. 5 (left). The calculated distribution of magnetic field in the coil and in the iron yoke at the coil current of 10 kA calculated with ANSYS in the 4L mirror configuration is shown in Fig. 5 (right).

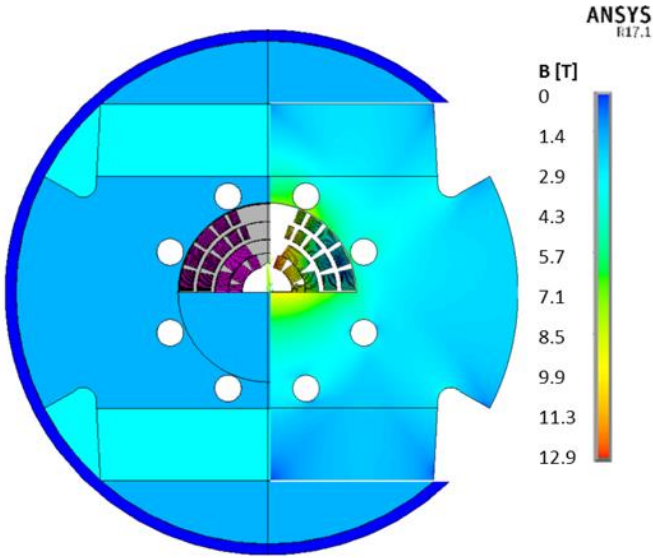

Fig. 5. SMCT coil in the 4L dipole mirror configuration with the MDPCT1 dipole inner coil and support structure (left) and calculated distribution of magnetic field induction in the coil and iron yoke (right).

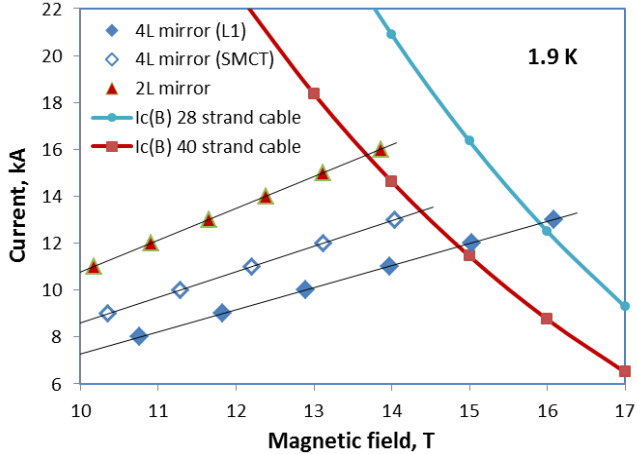

Fig. 6. $I_c(B)$ curves of 40-strand (SMCT coil) and 28-strand (inner coil) Nb$_3$Sn cables at 1.9 K, and load lines of SMCT and insert coils in 2L and 4L dipole mirror configurations.

The SMCT demo coil will be first connected to power supply (PS) and tested independently, and then in series with the MDPCT1 inner coil. In the first test SM structure will be tested with respect to the azimuthal Lorentz forces. Then, in the second test the radial Lorentz force component from the inner coil will be added. As it seen from Fig. 5 (right), the maximum field in the 4L mirror is in the inner-layer pole turn of the inner coil. The maximum field in the SMCT coil is lower and located in the coil inner-layer pole turn.

The $I_c(B)$ curves of the 40-strand (SMCT demo coil) and 28-strand (inner coil) Nb$_3$Sn cables and the load lines of SMCT and inner coils in the dipole mirror test configurations are shown in Fig. 6. The calculated conductor limit at 1.9 K of the individually powered SMCT coil in the mirror structure is of 13.7 T at 15.7 kA current. The conductor limit of the 4L mirror is determined by the inner coil and is 15.9 T at 12.9 kA current. The maximum field in the SMCT coil in the 4L mirror at the conductor limit is 13.9 T which is practically on the same level as in the 2L mirror.

TABLE I
MATERIAL PROPERTIES

| Structural element | Material | Thermal contract. (300-2 K), mm/m | Elasticity modulus, GPa |
|---|---|---|---|
| Coil (rad./azimuthal) | Nb$_3$Sn composite | 2.9/3.3 | 40/40 |
| Inner coil pole blocks | Ti-6Al-4V | 1.7 | 125 |
| Outer coil structure | Stainless steel | 2.9 | 215 |
| Inner coil wedges | Ti-6Al-4V | 3.2 | 120 |
| Coil-yoke spacer | Stainless steel | 2.9 | 210 |
| Yoke | Iron | 2.0 | 225 |

### B. Mirror Mechanical Analysis

A mechanical analysis of the SMCT demo coil in mirror configurations was performed using ANSYS to evaluate displacements of the key turns and stresses in the coil turns and SMCT coil structure. The key material properties used in the analysis are summarized in Table I.

The inner coil includes Ti-alloy poles and wedges. The outer coil is placed inside the stainless steel structure. The coil blocks are not glued to the SMCT coil structure to capture the effect of unloading under the Lorentz forces. In addition, each layer could slide with respect to the adjacent layers and to the iron yoke. The coils were pre-compressed during the assembly by placing the appropriate radial shims between the inner and outer coils and between the outer coil and the iron yoke.

Calculated distributions of the equivalent stress in coils after cool-down and with coil current of 15 kA in 2L and of 12 kA in 4L mirrors are shown in Fig. 7. Fig. 8 shows the calculated distributions of the equivalent stress in the outer coil structure and in the insert coil poles after cooling-down and with the appropriate coil currents in 2L and 4L mirror configurations. The maximum values (in MPa) of equivalent stress in both Nb$_3$Sn coils and coil structural elements are summarized in Table II.

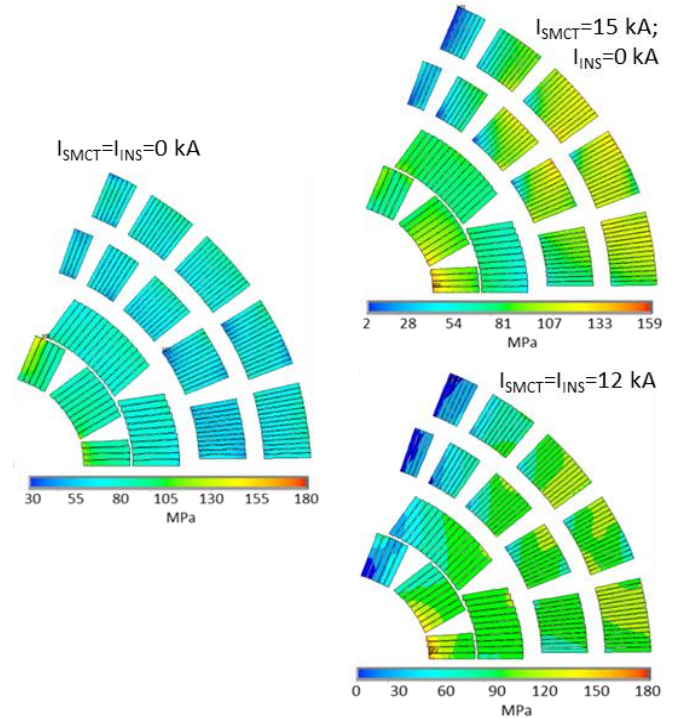

Fig. 7. The equivalent stress (in MPa) in the coils after cool-down without (left) and with current (right) in 2L (top) and 4L (bottom) mirrors.

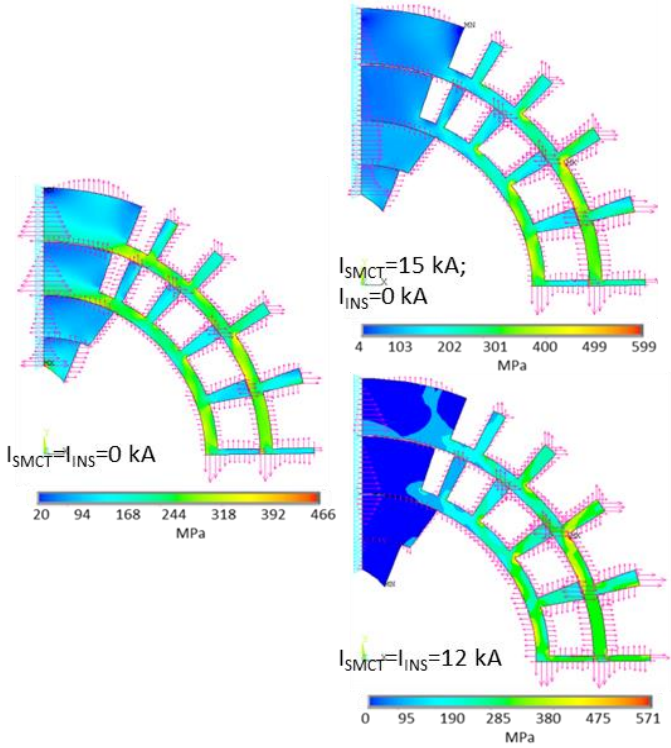

Fig. 8. Equivalent stress (in MPa) distribution in the SMCT coil structure and in the insert coil poles after cooling-down (left) and with current (right) in 2L (top) and 4L (bottom) mirrors.

TABLE II
MAX. EQUIVALENT STRESS IN $Nb_3Sn$ COILS AND COIL STRUCTURAL ELEMENTS IN DIPOLE MIRROR CONFIGURATION (MPA)

| Structural element | Material | 2L-mirror | | 4L-mirror | |
|---|---|---|---|---|---|
| | | 0 kA | 15 kA | 0 kA | 12 kA |
| Insert coil poles | Ti-6Al-4V | 466 | 178 | 463 | 168 |
| Insert coil | $Nb_3Sn$ | 180 | 159 | 180 | 180 |
| SMCT coil | $Nb_3Sn$ | 104 | 144 | 104 | 135 |
| SMCT structure | 316L | 464 | 599 | 464 | 571 |

The peak stress of 180 MPa after cool-down without current is in the pole block of layer 1 of the insert coil. With powering only SMCT demo coil to 15 kA (2L mirror) the peak stress reduces to 159 MPa and is in the layer 1 mid-plane block of the insert coil. With both coils connected in series (4L mirror) and powered to 12 kA the maximum stress is 180 MPa in the layer 1 mid-plane block of the insert coil.

In spite of the local peak stress of 466 MPa in the inner-layer of the insert coil, the stress in the SMCT structure after cool-down without current is less than ~350 MPa and in the insert poles less than 150 MPa. With powering only SMCT demo coil to 15 kA (2L mirror) the peak stress in the SMCT structure in average increases to ~450 MPa with the maximum stress point of 599 MPa in the corner of second mid-plane block of the outer layer. With both coils connected in series (4L mirror) and powered to 12 kA, the maximum stress in the SMCT coil structure is in the same place and on practically the same level of 571 MPa, whereas the average stress in the SMCT coil radial spacers less than 450 MPa. The maximum stress in the pole blocks of the insert coils is less than 200 MPa with average stress less than 100 MPa and 50 MPa in 2L and 4L mirrors respectively. All these stress numbers are acceptable for the coils in terms of $Nb_3Sn$ conductor degradation (<180 MPa) and with respect to the yield stress of coil structural materials (<600 MPa).

During powering the coil turns are shifted under Lorentz forces towards coil mid-plains. Absence of the reaction forces on the corresponding pole surfaces indicates that partial gaps can develop in the insert and SMCT demo coils at the currents close to the conductor limit. Analysis shows that these gaps can be up to 35-40 μm. The development of these gaps can lead to premature quenches and magnet training. These effects will be studied experimentally during mirror magnet training.

## IV. CONCLUSION

The SMCT coil is a new concept which was proposed at Fermilab for high-field and/or large-aperture accelerator magnets based on low-temperature and high-temperature superconductors [10], [11]. The SM structure is used to reduce large coil deformations under the Lorentz forces and, thus, the excessively high stresses in the coil and a separation of pole turns at high fields.

A 120-mm aperture $Nb_3Sn$ SMCT demo dipole coil with stress management has been developed and being fabricated at Fermilab to demonstrate and study the SMCT concept including coil design, fabrication technology and performance. The SMCT coil part design and coil fabrication technology were tested using plastic 3D printed parts and practice coil winding. The first set of SMCT coil parts for $Nb_3Sn$ dipole coils have been produced by GE Additive using DMLM technology. Printed parts have been measured using a laser scanning system with acceptable results. The SMCT demo coil winding is underway.

The first 120-mm aperture SMCT demo coil will be assembled and tested in a dipole mirror configuration. The mirror design and coil performance parameters have been developed and analyzed. The test of the mirror magnet with the SMCT demo coil is expected in the first part of 2022.

## ACKNOWLEDGMENTS

The authors thank Vadim Kashikhin for the magnetic optimization of the coil cross-section, Carry Lawless for the SMCT part procurement, Babatunde Oshinowo, Charles Wilson and Ted Beale for the parts measurements, Allen Rusy, James Karambis, Douglas Swanson, Michael Smego, Randall Wyatt, and Craig Bradford (Fermilab) for the technical support of this work.